\documentclass[11pt,a4paper]{article}
\usepackage[T1]{fontenc}
\usepackage[utf8]{inputenc}
\usepackage{lmodern,microtype}
\usepackage[margin=26mm]{geometry}
\usepackage{amsmath,amssymb,amsthm,mathtools}
\usepackage{booktabs,array,longtable,enumitem}
\usepackage{xcolor,xurl,hyperref}
\definecolor{linkblue}{RGB}{22,58,91}
\hypersetup{colorlinks=true,linkcolor=linkblue,citecolor=linkblue,urlcolor=linkblue,
 pdftitle={Equal Hamming Weights at Fixed Root Order and Applications to Constacyclic MDS Codes},pdfauthor={Li Zhu}}
\usepackage{fancyhdr}
\setlist[enumerate]{itemsep=3pt,topsep=5pt}
\newtheorem{theorem}{Theorem}[section]
\newtheorem{lemma}[theorem]{Lemma}
\newtheorem{proposition}[theorem]{Proposition}
\newtheorem{corollary}[theorem]{Corollary}
\theoremstyle{definition}

\newtheorem{example}[theorem]{Example}
\newtheorem{remark}[theorem]{Remark}
\newcommand{\F}{\mathbb F}
\newcommand{\Q}{\mathbb Q}
\newcommand{\Z}{\mathbb Z}

\newcommand{\EW}{\mathrm{EW}}
\DeclareMathOperator{\wt}{wt}
\DeclareMathOperator{\ord}{ord}
\DeclareMathOperator{\rad}{rad}

\DeclareMathOperator{\Res}{Res}

\DeclareMathOperator{\Gal}{Gal}
\DeclareMathOperator{\Min}{Min}
\newcommand{\Norm}[2]{N_{#1/#2}}

\numberwithin{equation}{section}

\title{Equal Hamming Weights at Fixed Root Order\\
and Applications to Constacyclic MDS Codes}
\author{Li Zhu$^{1}$ and Hongfeng Wu$^{2}$\footnote{Corresponding author.}
	\setcounter{footnote}{-1}
	\footnote{E-Mail addresses:
		lizhumath@pku.edu.cn (L. Zhu), whfmath@gmail.com (H. Wu)}
	\\[0.5ex]
	\small $^{1}$School of Mathematical Sciences, Guizhou Normal University, Guiyang, China
	\\
	\small $^{2}$College of Science, North China University of Technology, Beijing, China}
\date{}

\begin{document}
\maketitle
\begin{abstract}
For a fixed positive integer $n>1$, we study the prime powers $q$ coprime to $n$ for which all monic irreducible factors of $\Phi_n$ over $\F_q$ have the same coefficient Hamming weight. Cyclotomic orbit coefficients give a finite, effective classification of these fields. Outside finitely many computable characteristics, every factor has the support of its characteristic-zero orbit polynomial; the uniform condition $p>2^{\varphi(n)}$ suffices. In each remaining characteristic, extension exponents reduce to divisors of $\ord_n(p)$, and an integer variance identity expressed by polynomial gcd degrees decides equal weight. The same coefficient data determine the full weight distribution. As an application, we prove that an MDS constacyclic code with an irreducible generator forces every irreducible polynomial of the same root order to have full support, including factors associated with different constacyclic constants. This yields a uniform gcd and norm obstruction. We develop its short-multiplier extension, give a prime-order cubic family, and establish an explicit quadratic-residue exclusion without a prime-square alphabet or a near-square length hypothesis.
\end{abstract}



\section{Introduction}

The degrees of the irreducible factors of a cyclotomic polynomial over a finite field are determined by multiplicative orders. Their coefficient supports require additional information. Throughout the paper, the weight of a polynomial is the number of its nonzero coefficients. We fix $n>1$, write $q=p^a$ with $p\nmid n$, and consider the property
\[
 \EW_q(n):\qquad
 \text{all monic irreducible factors of $\Phi_n$ over $\F_q$ have equal weight}.
\]
The root order $n$ is fixed while both the characteristic $p$ and the extension exponent $a$ vary. This inverse problem admits a finite arithmetic description, even though a short uniform list of congruence classes need not result.

Let $G_n=(\Z/n\Z)^\times$. For a cyclic subgroup $H\le G_n$, the polynomial
\begin{equation}\label{eq:intro-orbit}
 F_H(X)=\prod_{h\in H}(X-\zeta_n^h)
\end{equation}
has algebraic integer coefficients in $K_H=\Q(\zeta_n)^H$. The support of $F_H$ is common to its characteristic-zero Galois conjugates. When $H=\langle q\bmod n\rangle$, reduction gives the irreducible factors in characteristic $p$. A nonzero coefficient can disappear only at a prime dividing its algebraic norm. Zero coefficients in characteristic zero remain zero under every reduction. These two cases must be kept separate.

Explicit Frobenius-orbit factorizations and their coefficient expressions belong to the established theory; see Graner~\cite{Graner} and the standard treatment in~\cite{LN}. Norm arguments transferring relations between complex and finite-field roots of unity occur in Duc--Leung--Schmidt~\cite{DLS}. A broader resultant formulation preserving finite families of evaluation relations appears in Breuillard--Varj\'u~\cite{BV}. Vanishing sums over finite fields were studied by Lam--Leung~\cite{LLfinite}. We use these principles to organize the simultaneous coefficient-zero distribution on a whole exact-order layer.

For the inverse problem, the resulting description has two parts. There is a finite set $\mathcal E_n$ of possible exceptional characteristics. For each $p\in\mathcal E_n$ with $p\nmid n$, there is a computable set $\mathcal A_{n,p}$ of divisors of $d_p=\ord_n(p)$ such that
\begin{equation}\label{eq:intro-class}
 \EW_{p^a}(n)
 \quad\Longleftrightarrow\quad
 \gcd(a,d_p)\in\mathcal A_{n,p}.
\end{equation}
All characteristics outside $\mathcal E_n$ satisfy equal weight for every exponent. Theorems~\ref{thm:generic} and~\ref{thm:classification} give the construction and proof. The set $\mathcal E_n$ is a set of possible exceptions, not a list of characteristics in which equal weight necessarily fails.

The exact finite-field test is an integer identity. Put $d=\ord_n(q)$, and let $U$ and $V$ be sums of degrees of gcds recording single and double coefficient-zero events. Theorem~\ref{thm:variance} proves
\begin{equation}\label{eq:intro-variance}
 \varphi(n)V-U^2
 =d^2\sum_{f<g}\bigl(\wt(f)-\wt(g)\bigr)^2.
\end{equation}
Here $f,g$ run through the irreducible factors of $\Phi_n$. Thus equal weight is decided by the vanishing of an ordinary nonnegative integer. An iterative gcd partition recovers the entire weight distribution.

Section~\ref{sec:mds} gives a separate application to constacyclic codes. We use ``irreducible generator'' explicitly, since the alternative convention of an irreducible check polynomial interchanges the relevant dimensions under duality. A code with an irreducible generator of degree $d$ can be MDS only if every polynomial of the same root order has weight $d+1$. The proof uses monomial equivalence across the allowed constacyclic constants. The coefficient norm test becomes a uniform exclusion tool, and a short-multiplier construction extends it to selected minors. The cyclotomic-minor norm construction is closely related to Xiang--Tang~\cite{XT}; we formulate and prove it in the orbit-coefficient notation used throughout this paper. The cubic application is also related to the classical study of Singer cyclic arcs~\cite{Maruta,Szonyi,MX}. An explicit higher-degree quadratic-residue family provides a further application.

\section{Orbit coefficients and finite-field factors}\label{sec:orbits}

Write $\Phi_n$ for the $n$th cyclotomic polynomial, choose a complex primitive $n$th root $\zeta_n$, and put
\[
 R_n=\Z[Z]/(\Phi_n(Z)),\qquad L_n=\Q(\zeta_n).
\]
We identify $G_n$ with $\Gal(L_n/\Q)$ by $\sigma_u(\zeta_n)=\zeta_n^u$. A bar over an integer polynomial denotes coefficient reduction modulo $p$. For $H\le G_n$ cyclic, let $d=|H|$ and define $C_{H,j}\in R_n$ by
\begin{equation}\label{eq:Cdef}
 \prod_{h\in H}(X-Z^h)
 =\sum_{j=0}^d C_{H,j}(Z)X^j.
\end{equation}
Whenever a polynomial representative is needed, $C_{H,j}(Z)$ denotes the unique integer remainder of degree less than $\varphi(n)$. Put
\[
 a_{H,j}=C_{H,j}(\zeta_n),\qquad
 J_H=\{j: a_{H,j}\ne0\},\qquad s_H=|J_H|.
\]
Both $0$ and $d$ belong to $J_H$. Indeed, $a_{H,d}=1$ and $a_{H,0}$ is a root of unity up to sign. All $a_{H,j}$ are algebraic integers in $K_H=L_n^H$.

\begin{lemma}\label{lem:factors}
Suppose $p\nmid n$ and $H=\langle q\bmod n\rangle$. Let $\xi$ be a primitive $n$th root in $\overline{\F}_p$. For $u\in G_n$, put
\[
 f_u(X)=\prod_{h\in H}(X-\xi^{uh}).
\]
Then the distinct $f_u$, indexed by $G_n/H$, are exactly the monic irreducible factors of $\Phi_n$ over $\F_q$. They have degree $d=|H|$, and
\[
 [X^j]f_u=\overline{C_{H,j}}(\xi^u).
\]
\end{lemma}
\begin{proof}
The $q$-Frobenius orbit of $\xi^u$ is $\{\xi^{uh}:h\in H\}$, because $u$ is a unit modulo $n$. Its length is $\ord_n(q)=d$. The orbit product is the minimal polynomial, and the coefficient formula follows from~\eqref{eq:Cdef}. Since $p\nmid n$, the roots of $\Phi_n$ are distinct and have exact order $n$.
\end{proof}

Define the weight enumerator of the exact-order layer by
\begin{equation}\label{eq:Wdef}
 W_{q,n}(T)=\sum_{\substack{f\mid\Phi_n\\ f\text{ monic irreducible over }\F_q}}T^{\wt(f)}.
\end{equation}
It has $\varphi(n)/d$ summands. Equal weight means that $W_{q,n}$ is a single monomial with this coefficient. This polynomial enumerates weights of defining polynomials, not weights of codewords in a linear code.

\begin{proposition}\label{prop:generic-support}
All Galois conjugates of $F_H$ in~\eqref{eq:intro-orbit} have support $J_H$.
\end{proposition}
\begin{proof}
For $u\in G_n$, the automorphism $\sigma_u(\zeta_n)=\zeta_n^u$ carries $a_{H,j}$ to the coefficient of $X^j$ in $\prod_{h\in H}(X-\zeta_n^{uh})$. A field automorphism preserves, in both directions, whether an element is zero.
\end{proof}

\section{Norm specialization and generic support}\label{sec:norms}

Define the positive integer
\begin{equation}\label{eq:Bdef}
 B(n,H)=\prod_{j\in J_H}\left|\Norm{K_H}{\Q}(a_{H,j})\right|.
\end{equation}
The endpoint coefficients contribute $1$. The product omits characteristic-zero zero coefficients. This convention differs from the full-support obstruction introduced in Section~\ref{sec:mds}.

\begin{lemma}\label{lem:norm-reduction}
Let $p\nmid n$. For any nonzero $\beta\in\mathcal O_{K_H}$,
\[
 p\mid\Norm{K_H}{\Q}(\beta)
 \quad\Longleftrightarrow\quad
 \beta\in\mathfrak p\text{ for some prime ideal }\mathfrak p\mid p\text{ of }K_H.
\]
If $H=\langle p^a\bmod n\rangle$, reduction of the orbit products, over all primes and residue-field embeddings, gives precisely the factors in Lemma~\ref{lem:factors}.
\end{lemma}
\begin{proof}
The factorization of the principal ideal $(\beta)$ shows that its norm is divisible by $p$ exactly when a prime above $p$ divides $(\beta)$. Alternatively, this follows from the determinant of multiplication by $\beta$ on $\mathcal O_{K_H}$.

For the second assertion, extend a prime of $K_H$ above $p$ to $L_n$. The reduction of $\zeta_n$ has exact order $n$: the factorization $X^n-1=\prod_{e\mid n}\Phi_e$ is separable in characteristic $p$, so different divisor layers have disjoint roots. Each residue image is therefore a primitive $n$th root. The $p^a$-orbit of a residue root $\xi$ is $\{\xi^h:h\in H\}$, and its orbit product is its minimal polynomial over $\F_{p^a}$. Conversely, fix one residue map $\rho$ with $\rho(\zeta_n)=\xi$. For $u\in G_n$, the maps $\rho\circ\sigma_u$ send $\zeta_n$ to $\xi^u$ and cover every primitive root. Thus all factors occur.
\end{proof}

\begin{theorem}\label{thm:generic}
Let $q=p^a$, $p\nmid n$, and $H=\langle q\bmod n\rangle$, with $d=|H|$.
\begin{enumerate}[label=\textup{(\roman*)}]
\item Every irreducible factor of $\Phi_n$ over $\F_q$ has support contained in $J_H$.
\item All factors have support exactly $J_H$ if and only if $p\nmid B(n,H)$. In particular,
\[
 p\nmid B(n,H)\quad\Longrightarrow\quad
 W_{q,n}(T)=\frac{\varphi(n)}dT^{s_H}.
\]
\item For every $j\in J_H$,
\begin{equation}\label{eq:norm-bound}
 1\le\left|\Norm{K_H}{\Q}(a_{H,j})\right|
 \le \binom{d}{j}^{\varphi(n)/d}\le 2^{\varphi(n)}.
\end{equation}
Consequently, $p>2^{\varphi(n)}$ implies $\EW_{p^a}(n)$ for every $a\ge1$.
\end{enumerate}
\end{theorem}
\begin{proof}
A zero element of $R_n$ specializes to zero at every primitive root, proving (i). If a nonzero coefficient disappears in some factor, Lemma~\ref{lem:norm-reduction} implies that its norm is divisible by $p$. Conversely, if $p$ divides one of the norms, choose a prime and a residue embedding at which that coefficient vanishes. The corresponding factor has lost a coefficient from $J_H$. This proves (ii), including its ``only if'' assertion.

Each complex conjugate of $a_{H,j}$ is a sum, up to sign, of $\binom{d}{j}$ roots of unity. Its absolute value is at most $\binom{d}{j}$. Since $[K_H:\Q]=\varphi(n)/d$, multiplication over the embeddings gives~\eqref{eq:norm-bound}. The norm is a nonzero integer. A prime larger than $2^{\varphi(n)}$ divides none of these individual norms, for any cyclic subgroup $H$.
\end{proof}

\begin{remark}
The condition $p\mid B(n,H)$ detects failure of generic support in at least one factor. It does not imply failure of equal weight: all factors can lose the same number of coefficients. Also, generic support need not be full support when $n$ is composite.
\end{remark}

The norm test has an entirely integral resultant version. For $j\in J_H$, put
\begin{equation}\label{eq:resultant-norm}
 R_{H,j}=\Res_Z(\Phi_n(Z),C_{H,j}(Z)).
\end{equation}
Then
\[
 R_{H,j}=\Norm{L_n}{\Q}(a_{H,j})
 =\Norm{K_H}{\Q}(a_{H,j})^d.
\]
Thus the fixed-field norm and the resultant have identical sets of prime divisors. This also permits a computation in integer polynomial arithmetic without constructing $K_H$ explicitly.

\begin{proposition}\label{prop:prime-nonzero}
If $n=\ell$ is prime, then $J_H=\{0,1,\ldots,d\}$ for every subgroup $H\le G_\ell$.
\end{proposition}
\begin{proof}
For an interior coefficient, reduce the exponents in its elementary symmetric sum modulo $\ell$. Up to a global sign this gives a polynomial $E_j(Z)\in\Z[Z]$ of degree at most $\ell-1$, whose coefficients are nonnegative and whose value at $1$ is $\binom{d}{j}$. If $E_j(\zeta_\ell)=0$, then $E_j=c\Phi_\ell$ for an integer $c$. Hence $\ell\mid \binom{d}{j}$. This is impossible for $0\le j\le d<\ell$, since all factorials involved are prime to $\ell$. The endpoint coefficients are already nonzero.
\end{proof}

\section{Gcd moments and the full weight distribution}\label{sec:gcd}

Fix $q=p^a$ and $H=\langle q\bmod n\rangle$, and write $d=|H|$. For $1\le j\le d-1$ and $1\le i<j\le d-1$, define ordinary nonnegative integers
\begin{align}
 \nu_j&=\deg\gcd(\overline{\Phi_n},\overline{C_{H,j}}),\label{eq:nu1}\\
 \nu_{ij}&=\deg\gcd(\overline{\Phi_n},\overline{C_{H,i}},\overline{C_{H,j}}),\label{eq:nu2}\\
 U&=\sum_{j=1}^{d-1}\nu_j,\qquad
 V=U+2\sum_{1\le i<j\le d-1}\nu_{ij}.\label{eq:UV}
\end{align}
Every gcd here is computed in $\F_p[Z]$. Empty sums, including the case $d=1$, are zero.

\begin{theorem}\label{thm:variance}
Let $\mathcal F_{q,n}$ be the set of monic irreducible factors of $\Phi_n$ over $\F_q$. Then, with the sum taken over unordered pairs of distinct factors,
\begin{equation}\label{eq:variance}
 \Delta_{q,n}:=\varphi(n)V-U^2
 =d^2\sum_{\{f,g\}\subseteq\mathcal F_{q,n}}
       \bigl(\wt(f)-\wt(g)\bigr)^2\ge0.
\end{equation}
In particular,
\[
 \EW_q(n)\quad\Longleftrightarrow\quad\Delta_{q,n}=0.
\]
When equal weight holds, the common weight is
\begin{equation}\label{eq:common-weight}
 d+1-\frac{U}{\varphi(n)}.
\end{equation}
All irreducible factors have full support if and only if $U=0$.
\end{theorem}
\begin{proof}
For $f\in\mathcal F_{q,n}$, let $z_f$ be the number of its zero interior coefficients. Then $\wt(f)=d+1-z_f$. The gcd in~\eqref{eq:nu1} counts primitive roots whose associated factor has a zero $j$th coefficient. Each factor contributes its $d$ roots. A double gcd counts simultaneous zeros in the same way. Therefore
\[
 U=d\sum_f z_f,\qquad V=d\sum_f z_f^2.
\]
The number of factors is $m=\varphi(n)/d$. The elementary identity
\[
 m\sum_f z_f^2-\Bigl(\sum_f z_f\Bigr)^2
 =\sum_{\{f,g\}}(z_f-z_g)^2
\]
proves~\eqref{eq:variance}. If it vanishes, all $z_f$ coincide and equal $U/\varphi(n)$, giving~\eqref{eq:common-weight}. Finally, $U=0$ is equivalent to every $z_f=0$.
\end{proof}

\begin{remark}
The equality $\Delta_{q,n}=0$ is tested in $\Z$, not in $\F_p$. Reducing the integer variance modulo $p$ would lose the equivalence with equal weight. The theorem uses $O(d^2)$ gcd computations, but the construction of the orbit coefficients and the degrees of their representatives remain part of the cost. No polynomial-time bound in $\log n$, or general advantage over factorization, is asserted.
\end{remark}

\begin{proposition}\label{prop:distribution}
The full polynomial $W_{q,n}(T)$ can be computed by an iterative gcd partition over $\F_p$, without first recovering all individual factors over $\F_q$.
\end{proposition}
\begin{proof}
Initially set $D_0=\overline{\Phi_n}$ and all other buckets equal to $1$. After processing coefficients $1,\ldots,j-1$, the roots of $D_z$ are exactly the primitive roots with $z$ zeros among those coefficients. For coefficient $j$, split each nontrivial bucket into
\[
 E_z=\gcd(D_z,\overline{C_{H,j}}),\qquad D_z/E_z.
\]
Place $E_z$ in the new bucket $z+1$ and $D_z/E_z$ in the new bucket $z$, multiplying contributions to the same bucket. Separability makes these products disjoint root partitions. At termination, each $D_z$ is a union of $q$-orbits and
\begin{equation}\label{eq:bucket-W}
 W_{q,n}(T)=\sum_{z=0}^{d-1}\frac{\deg D_z}{d}\,T^{d+1-z}.
\end{equation}
There are at most $j$ nonempty buckets before step $j$, so at most $O(d^2)$ gcd operations occur. Quotients and products must also be computed.
\end{proof}

\section{A finite inverse classification}\label{sec:classification}

\begin{lemma}\label{lem:exponent-reduction}
Fix $p\nmid n$, put $d_p=\ord_n(p)$, and let $a\ge1$. If $g=\gcd(a,d_p)$, then
\begin{equation}\label{eq:exponent-reduction}
 W_{p^a,n}(T)=W_{p^g,n}(T).
\end{equation}
\end{lemma}
\begin{proof}
In the cyclic group $\langle p\bmod n\rangle$ of order $d_p$, the elements $p^a$ and $p^g$ generate the same subgroup. Thus the Frobenius orbit partitions of the primitive $n$th roots agree. Their orbit polynomials have coefficients in $\F_{p^g}\subseteq\F_{p^a}$, and give the same lists of weights in both fields.
\end{proof}

For a cyclic subgroup $H\le G_n$, retain the positive integer $B(n,H)$ from~\eqref{eq:Bdef}, and set
\begin{equation}\label{eq:En}
 \mathcal E_n=\{p\text{ prime}:p\nmid n,
       \ p\mid\prod_{\substack{H\le G_n\\H\text{ cyclic}}} B(n,H)\}.
\end{equation}
For $p\in\mathcal E_n$, define
\begin{equation}\label{eq:Anp}
 \mathcal A_{n,p}=
 \{g\mid d_p:\Delta_{p^g,n}=0\}.
\end{equation}

\begin{theorem}\label{thm:classification}
For every fixed $n>1$, the set of prime powers $q=p^a$ with $p\nmid n$ satisfying $\EW_q(n)$ has the following finite effective description:
\begin{enumerate}[label=\textup{(\roman*)}]
\item if $p\notin\mathcal E_n$, every exponent $a\ge1$ is allowed;
\item if $p\in\mathcal E_n$, exactly the exponents satisfying
\[
 \gcd(a,d_p)\in\mathcal A_{n,p}
\]
are allowed.
\end{enumerate}
The set $\mathcal E_n$ and every set $\mathcal A_{n,p}$ can be obtained by finite integer polynomial arithmetic, integer factorization, and polynomial gcd computations over prime fields. Every prime in $\mathcal E_n$ is at most $2^{\varphi(n)}$.
\end{theorem}
\begin{proof}
There are finitely many cyclic subgroups of $G_n$. Formula~\eqref{eq:Cdef} gives their coefficient representatives, and integer resultants~\eqref{eq:resultant-norm} compute the prime divisors of their nonzero coefficient norms. Thus~\eqref{eq:En} is effective and finite. The bound on its primes follows from~\eqref{eq:norm-bound}.

If $p\notin\mathcal E_n$, Theorem~\ref{thm:generic} applies to $H=\langle p^a\rangle$ for every $a$, proving (i). For $p\in\mathcal E_n$, there are only finitely many divisors $g$ of $d_p$. Compute $\Delta_{p^g,n}$ by Theorem~\ref{thm:variance}, and use Lemma~\ref{lem:exponent-reduction}. This proves (ii).
\end{proof}

The theorem classifies all fields for a fixed root order by finite data. It does not interchange the quantifiers to classify all root orders for a fixed field. The integer products used to locate possible exceptions can also be substantially larger than the set of actual exceptions.

There is a useful alternative after factoring $\Phi_n$ over the prime field. Suppose
\[
 \overline{\Phi_n}=\prod_{i=1}^{\varphi(n)/d_p}P_i,
 \qquad \deg P_i=d_p.
\]
For $g\mid d_p$, put $H_g=\langle p^g\bmod n\rangle$ and $d=d_p/g$, and define
\[
 w_i(g)=d+1-
 \#\{1\le j\le d-1:P_i\mid\overline{C_{H_g,j}}\}.
\]

\begin{proposition}\label{prop:prime-factor-formula}
For $g=\gcd(a,d_p)$,
\begin{equation}\label{eq:prime-factor-W}
 W_{p^a,n}(T)=g\sum_{i=1}^{\varphi(n)/d_p}T^{w_i(g)}.
\end{equation}
\end{proposition}
\begin{proof}
The $p$-orbit corresponding to $P_i$ splits into $g$ orbits under $p^g$. The coefficient $p$-Frobenius permutes the resulting factors transitively and preserves their zero coefficients. For a representative root $\alpha$ of $P_i$, an interior coefficient vanishes exactly when $C_{H_g,j}(\alpha)=0$. Since $C_{H_g,j}$ is defined over $\F_p$, this is equivalent to $P_i\mid\overline{C_{H_g,j}}$. Each of the $g$ factors therefore has weight $w_i(g)$. Now apply~\eqref{eq:exponent-reduction}.
\end{proof}

\section{Symmetries and elementary cases}\label{sec:symmetry}

\begin{proposition}\label{prop:symmetries}
Let $q=p^a$, $p\nmid n$. The factor-weight function on $G_n/\langle q\rangle$ is invariant under multiplication by $p$ and by $-1$. Consequently the number of distinct weights is at most
\[
 [G_n:\langle p,-1\rangle].
\]
If $\langle p,-1\rangle=G_n$, then $\EW_{p^a}(n)$ holds for every $a\ge1$.
\end{proposition}
\begin{proof}
Applying coefficient $p$-Frobenius sends the factor with root $\xi^u$ to that with root $\xi^{pu}$ and preserves zero coefficients. Taking the monic reciprocal sends $u$ to $-u$ and preserves weight. Since $\langle q\rangle\subseteq\langle p,-1\rangle$, the orbit bound follows.
\end{proof}

\begin{lemma}\label{lem:degree-two}
If $\ord_n(q)=2$, then $\EW_q(n)$ holds.
\end{lemma}
\begin{proof}
An irreducible quadratic with root $\alpha$ is
\[
 X^2-(\alpha+\alpha^q)X+\alpha^{q+1}.
\]
Only its middle coefficient can vanish. In characteristic two, vanishing would make the polynomial inseparable and is impossible. In odd characteristic, it vanishes exactly when $n$ is even and $q-1\equiv n/2\pmod n$: the equation is $\alpha^{q-1}=-1=\alpha^{n/2}$. This condition is independent of the primitive root $\alpha$, since every unit modulo an even $n$ is odd. Thus all factors have weight two, or all have weight three.
\end{proof}

\begin{proposition}\label{prop:small-group}
If $G_n$ is cyclic of order $2\ell$, where $\ell$ is an odd prime, then $\EW_q(n)$ holds for every $q$ coprime to $n$. The same conclusion holds whenever $\varphi(n)\le6$.
\end{proposition}
\begin{proof}
In the first assertion, $d=\ord_n(q)$ is one of $1,2,\ell,2\ell$. Degrees one and two are covered by linear factors and Lemma~\ref{lem:degree-two}; degree $2\ell$ gives one factor. If $d=\ell$, there are two factors and $-1\notin\langle q\rangle$, so the reciprocal exchanges them. For $\varphi(n)\le6$, the possible values of $\varphi(n)$ are $1,2,4,6$. For values at most four, the only cases besides one factor or linear factors have degree two. For value six the unit group is cyclic, and the first assertion applies.
\end{proof}

\begin{example}\label{ex:n7}
At root order $7$, the subgroup of order three has orbit coefficients in $\Q(\sqrt{-7})$. The two interior coefficients have absolute norm $2$. In characteristic two,
\[
 \Phi_7=(X^3+X+1)(X^3+X^2+1),
\]
so both cubic factors have weight three instead of their generic weight four. Thus $2\mid B(7,H)$, but equal weight persists. Proposition~\ref{prop:small-group} in fact proves equal weight at root order seven over every allowed field.
\end{example}

\begin{example}\label{ex:n13}
In characteristic three,
\begin{align*}
 \Phi_{13}={}&(X^3-X-1)(X^3+X^2-1)\\
 &\cdot(X^3+X^2+X-1)(X^3-X^2-X-1).
\end{align*}
Since $\ord_{13}(3)=3$, all four displayed factors are irreducible. Therefore
\[
 W_{3^a,13}(T)=
 \begin{cases}
 2T^3+2T^4,&3\nmid a,\\
 12T^2,&3\mid a.
 \end{cases}
\]
This is an exact inverse classification within characteristic three, illustrating the divisor condition on $\gcd(a,d_p)$ rather than a search over extension degrees.
\end{example}

\section{Necessary conditions for constacyclic MDS codes}\label{sec:mds}

Throughout this section, $N$ denotes code length, and $n$ denotes multiplicative root order. Let $\lambda\in\F_q^\times$ and let $f\mid X^N-\lambda$ be monic irreducible of degree $d<N$. Put
\[
 C_f=\langle f\rangle\subseteq R_\lambda:=\F_q[X]/(X^N-\lambda).
\]
The code has dimension $N-d$, so its MDS distance would be $d+1$. An irreducible check polynomial convention instead gives a code of dimension $d$; its dual has a reciprocal irreducible generator. Root order and polynomial weight are preserved by reciprocation, so the conclusions below apply to that convention as well.

Cyclotomic reduction of minors and norm obstructions are used in Xiang--Tang~\cite[Theorems 6--7]{XT}. The three-dimensional geometric setting is studied in~\cite{Maruta,Szonyi,MX}, and isometries between constacyclic ambient rings are part of the classical framework of~\cite{CFL}. We give the needed statements and proofs here in the orbit-coefficient notation of Section~\ref{sec:orbits}.

\subsection{Root-order equivalence and maximal equal weight}

\begin{lemma}\label{lem:const-equivalence}
Let $\alpha$ be a root of $f$ of order $n$, so $\alpha^N=\lambda$. For $u\in G_n$, let $f_u=\Min_{\F_q}(\alpha^u)$. Then
\[
 \langle f_u\rangle\subseteq R_{\lambda^u}
 \quad\text{and}\quad C_f\subseteq R_\lambda
\]
are monomially equivalent.
\end{lemma}
\begin{proof}
Choose a positive integer $b\equiv u\pmod n$ with $\gcd(b,N)=1$. Such a choice is possible by imposing $b\equiv1\pmod\ell$ for the primes $\ell\mid N$ not dividing $n$, and applying the Chinese remainder theorem. Primes dividing both $N$ and $n$ already do not divide $b$.

The map
\[
 \Psi_b:R_{\lambda^u}\longrightarrow R_\lambda,
 \qquad c(X)\longmapsto c(X^b)
\]
is well-defined because $(X^b)^N=\lambda^b=\lambda^u$. It maps the coordinate basis by
\[
 X^i\longmapsto\lambda^{\lfloor bi/N\rfloor}X^{bi\bmod N}.
\]
Thus it is a monomial linear bijection and an algebra isomorphism. Since $f_u(\alpha^b)=0$, evaluation at $\alpha$ shows that $\Psi_b(\langle f_u\rangle)\subseteq\langle f\rangle$. Both ideals have dimension $N-d$, so they are equal.
\end{proof}

\begin{theorem}\label{thm:mds-full}
With the preceding notation,
\begin{equation}\label{eq:distance-bound}
 d_{\min}(C_f)\le
 \min_{\substack{g\mid\Phi_n\\g\text{ monic irreducible over }\F_q}}\wt(g).
\end{equation}
In particular, if $C_f$ is MDS, then
\begin{equation}\label{eq:mds-W}
 W_{q,n}(T)=\frac{\varphi(n)}dT^{d+1}.
\end{equation}
Failure of this equality excludes every length-$N$ irreducible-generator constacyclic code of root order $n$, across all its allowed constacyclic constants.
\end{theorem}
\begin{proof}
Every such $g$ equals some $f_u$ in Lemma~\ref{lem:const-equivalence}. It is a nonzero codeword in its own length-$N$ code because $\deg g=d<N$. Monomial equivalence preserves distance, proving~\eqref{eq:distance-bound}. Every degree-$d$ polynomial has weight at most $d+1$, whereas an MDS code $C_f$ must have distance $d+1$. Thus all the factors have weight $d+1$.
\end{proof}

\begin{proposition}\label{prop:projective-period}
Put $s=n/\gcd(n,q-1)$. Then $s\mid N$. If $s<N$, then $d_{\min}(C_f)=2$. Consequently, for $d\ge2$,
\begin{equation}\label{eq:period-condition}
 C_f\text{ MDS}\quad\Longrightarrow\quad N=s.
\end{equation}
Writing $t=\ord(\lambda)$, the equality $N=s$ is equivalent to
\begin{equation}\label{eq:period-parameter}
 n=Nt,\qquad\gcd\left(N,\frac{q-1}{t}\right)=1.
\end{equation}
In particular, $p\mid N$ excludes the MDS property for $d\ge2$.
\end{proposition}
\begin{proof}
The order of $\alpha\F_q^\times$ in $\F_{q^d}^\times/\F_q^\times$ is $s$. Since $\alpha^N\in\F_q^\times$, it divides $N$. If $s<N$, the polynomial $X^s-\alpha^s$ is a weight-two codeword divisible by $f$. No monomial vanishes at the nonzero root $\alpha$, so distance one is impossible.

If $s=N$, write $c=\gcd(n,q-1)$; then $n=Nc$ and $t=n/\gcd(n,N)=c$. Hence $\gcd(N,(q-1)/t)=1$. Conversely,~\eqref{eq:period-parameter} gives $\gcd(n,q-1)=t$, so $s=N$. Finally $p\nmid n$, hence $p\nmid s$.
\end{proof}

For $d=1$, $C_f$ is the usual $[N,N-1,2]$ code and is MDS; this explains the stated degree restriction in~\eqref{eq:period-condition}. For $d=2$, the projective-period condition is sufficient as well as necessary. Indeed, in an $\F_q$-basis of $\F_{q^2}$ the columns $1,\alpha,\ldots,\alpha^{N-1}$ form a parity-check matrix. If $s=N$, no two columns are proportional, so the distance is three.

\subsection{Coefficient gcds and norms as uniform obstructions}

Define a second norm product, now including every interior position:
\begin{equation}\label{eq:Bzero}
 B_0(n,H)=\prod_{j=1}^{d-1}
 \left|\Norm{K_H}{\Q}(a_{H,j})\right|\in\Z_{\ge0}.
\end{equation}
If an interior coefficient is zero in characteristic zero, then $B_0(n,H)=0$. We use the usual convention that every prime divides zero. This is the distinction from the positive product $B(n,H)$ in Section~\ref{sec:norms}.

\begin{theorem}\label{thm:mds-arithmetic}
Let $H=\langle q\bmod n\rangle$. The following conditions are equivalent:
\begin{enumerate}[label=\textup{(\roman*)}]
\item every irreducible factor of $\Phi_n$ over $\F_q$ has full support;
\item $U=0$ in~\eqref{eq:UV};
\item $\gcd(\overline{\Phi_n},\overline{C_{H,j}})=1$ for every $1\le j\le d-1$;
\item $p\nmid B_0(n,H)$.
\end{enumerate}
Each is necessary for $C_f$ to be MDS. If any condition fails, the entire same-root-order, length-$N$ family is excluded.
\end{theorem}
\begin{proof}
The equivalence of (i)--(iii) is Theorem~\ref{thm:variance} and the nonnegativity of each gcd degree. If an interior characteristic-zero coefficient is zero, all its conjugates specialize to zero and $B_0=0$. Otherwise Lemma~\ref{lem:norm-reduction}, applied to each coefficient, proves the equivalence with (iv). The final assertion follows from Theorem~\ref{thm:mds-full}.
\end{proof}

In particular, a norm prime may preserve equal weight while ruling out MDS: Example~\ref{ex:n7} has exactly this behavior. A variance test alone would miss such an exclusion.

\begin{corollary}\label{cor:sparse}
If a same-order layer is equal-weight with common weight $w\le d$, all its length-$N$ codes as above are non-MDS. Thus common three-term layers are excluded for $d\ge3$, and common four-term layers for $d\ge4$. If $f(X)=G(X^h)$ with $h>1$, then $C_f$ is non-MDS.
\end{corollary}
\begin{proof}
The first assertion is~\eqref{eq:distance-bound}. In the last assertion $h\mid d$ and
\[
 \wt(f)=\wt(G)\le d/h+1\le d.
\]
\end{proof}

After the projective-period filter, candidates with $d\ge2$ have $p\nmid N$ and are simple-root codes. Support-compression exclusions in this setting are already reflected in the arithmetic Singleton bound (ASB) of Zhu--Wu~\cite[Theorem 4.6]{ASB}. If the root order is squarefree, that theorem gives ASB equal to $d+1$, the ordinary Singleton bound. In this range Theorem~\ref{thm:mds-arithmetic} can supply a strict additional restriction by detecting characteristic-dependent coefficient cancellation.

\begin{lemma}\label{lem:scalar}
Suppose $\gcd(N,q-1)=1$. Then every $\lambda\in\F_q^\times$ has a unique $N$th root $c\in\F_q^\times$, and variable scaling gives a monomial equivalence from cyclic codes of length $N$ to $\lambda$-constacyclic codes. It preserves degrees, irreducibility, and coefficient supports of the scaled generators.
\end{lemma}
\begin{proof}
The map $c\mapsto c^N$ is an automorphism of the cyclic group $\F_q^\times$. Substitution $X\mapsto X/c$ identifies the relations $X^N-1$ and $X^N-\lambda$ up to a nonzero scalar. On coordinates it multiplies the $i$th entry by $c^{-i}$. A monic degree-$d$ generator is transformed to $c^df(X/c)$.
\end{proof}

\begin{proposition}\label{prop:prime-ASB}
Let $N=\ell$ be prime and $d=\ord_\ell(q)>1$. For every
$\lambda\in\F_q^\times$, the ASB of a code generated by an irreducible
degree-$d$ factor of $X^\ell-\lambda$ is $d+1$, equal to Singleton.
\end{proposition}
\begin{proof}
By Lemma~\ref{lem:scalar}, write $c^\ell=\lambda$ with $c\in\F_q^\times$.
A root of a degree-$d$ factor is $c\xi$, where $\xi$ is primitive of
order $\ell$. Put $t=\ord(c)\mid q-1$. Since $\gcd(\ell,t)=1$, its order
is $n=\ell t$. In the notation of Zhu--Wu~\cite[Theorem 4.6]{ASB},
$\omega=\ord_{\rad(n)}(q)=d$: every prime divisor of $t$ has order one.
The possible $2$-adic correction does not occur, because $8\mid n$ would
imply $8\mid t$ and $q\equiv1\pmod8$. Every prime-power divisor of $t$
already divides $q-1$, and the exponent of $\ell$ in $n$ is one. Thus all
lifting exponents in that formula are zero and ASB is $\omega+1=d+1$.
\end{proof}

\subsection{Short multipliers and orbit-coefficient minors}\label{subsec:windows}

Put $k=N-d$. For a degree-$d$ polynomial $f=\sum_{i=0}^d a_iX^i$ with $a_0a_d\ne0$, and $0\le m\le k-1$, define the convolution matrix
\begin{equation}\label{eq:Tmatrix}
 T_m(f)=(a_{i-j})_{\substack{0\le i\le d+m\\0\le j\le m}},
 \qquad a_t=0\quad(t<0\text{ or }t>d).
\end{equation}
It maps the coefficients of a multiplier of degree at most $m$ to those of its product with $f$.

\begin{proposition}\label{prop:windows}
The code $C_f$ is MDS if and only if, for every $0\le m\le k-1$, every square submatrix formed by $m+1$ rows of $T_m(f)$ is nonsingular. It suffices to impose this for $m=k-1$. For $m=0$, it is precisely the full-support condition on $f$.
\end{proposition}
\begin{proof}
If a stated submatrix is singular, there is a nonzero polynomial $A$ of degree at most $m$ such that $Af$ vanishes at $m+1$ selected coefficient positions among $0,\ldots,d+m$. Thus
\[
 0\ne Af\in C_f,\qquad\wt(Af)\le(d+m+1)-(m+1)=d.
\]
Its degree is less than $N$, so no reduction of the product is involved. Conversely, a nonzero codeword is uniquely $Af$ with $\deg A=m\le k-1$. If its weight is at most $d$, its coefficient vector among these $d+m+1$ positions has at least $m+1$ zeros, producing a singular row submatrix. The top value $m=k-1$ is the usual maximal-minor criterion for a generator matrix, so it suffices on its own.
\end{proof}

Here take $H=\langle q\bmod n\rangle$, with $|H|=d$. Apply this construction to $F_H$ over $K_H$, and let $D_{m,I}$ be the determinant for a set $I$ of $m+1$ rows. These determinants are algebraic integers expressed in the orbit coefficients. Define
\[
 B_m(n,H;N)=\prod_{|I|=m+1}
       \left|\Norm{K_H}{\Q}(D_{m,I})\right|,
\]
again allowing zero products. In particular $B_0(n,H;N)=B_0(n,H)$, since the endpoint factors have norm one.

\begin{proposition}\label{prop:minor-norm}
Under the code hypotheses of this section, MDS implies $p\nmid B_m(n,H;N)$ for every $0\le m\le k-1$. Moreover,
\[
 C_f\text{ MDS}\quad\Longleftrightarrow\quad
 p\nmid B_{k-1}(n,H;N).
\]
\end{proposition}
\begin{proof}
Fix a residue map $\rho$ on $\Z[\zeta_n]$ for which
$\rho(\zeta_n)=\alpha$, and let $\mathfrak P_0$ be its prime above $p$.
If a determinant is zero in characteristic zero, it vanishes under every
reduction. Otherwise, divisibility of its norm by $p$ gives a prime
$\mathfrak P\mid p$ of $L_n$ containing it. The Galois group acts transitively
on the primes above $p$, so some $\sigma\in\Gal(L_n/\Q)$ carries
$\mathfrak P$ to $\mathfrak P_0$. The conjugate determinant $\sigma D_{m,I}$
is the same row minor of $T_m(\sigma F_H)$, and vanishes under $\rho$.
The polynomial $\rho(\sigma F_H)$ is one of the factors $f_u$.
Proposition~\ref{prop:windows} excludes its MDS property, and
Lemma~\ref{lem:const-equivalence} transfers that conclusion to $C_f$.
Conversely, a non-MDS $C_f$ has a vanishing top minor by
Proposition~\ref{prop:windows}. It is a reduction of the corresponding
orbit-coefficient determinant (or its Galois conjugate), whose norm is
therefore divisible by $p$. Norms are invariant under Galois conjugation,
which proves the stated equivalence.
\end{proof}

This is a self-contained version of the cyclotomic-minor reduction principle used in~\cite{XT}. The coefficient test is the first member of this hierarchy. Its smaller data set makes it a natural preliminary arithmetic filter; the full top-level product can be much larger. Avoiding a selected lower-level product alone does not establish MDS.

\subsection{A prime-length cubic family}
\label{sec:prime-cubic-family}

This section records a full-support obstruction for irreducible-generator
constacyclic codes of prime length.  The obstruction involves one fixed
cyclotomic integer norm.  It is a sufficient obstruction to the MDS property;
its complement is not asserted to be an MDS criterion.

The projective interpretation belongs to the established theory of cyclic arcs
and Singer orbits, including the earlier work of Maruta and Sz\H{o}nyi as discussed in Momihara--Xiang~\cite[Remark 3.2]{MX}.
The examples below serve as explicit arithmetic certificates and as a comparison
between generator weights and the minimum distance of their codes.

Fix henceforth a prime $\ell\equiv1\pmod3$, let $\zeta=\zeta_\ell$ be a complex primitive $\ell$th root, and let
\[
 H=\{1,u,u^2\}\subseteq(\mathbb Z/\ell\mathbb Z)^\times
\]
be the unique subgroup of order three. Choose $1\le u<\ell$ representing a generator of $H$ and set
\[
 L=\mathbb Q(\zeta),\qquad K=L^H,\qquad
 \eta_H=\sum_{h\in H}\zeta^h,\qquad
 B_\ell=\left|N_{K/\mathbb Q}(\eta_H)\right|.
\]
The number $B_\ell$ depends on $\ell$, not on the chosen primitive root or the
choice between the two generators of $H$.

\begin{proposition}
\label{prop:prime-cubic-norm}
The integer $B_\ell$ is positive.  The orbit polynomial is
\[
 F_H(X)=X^3-\eta_HX^2+\eta_{-H}X-1,
\]
and the absolute norms of its two internal coefficients are both $B_\ell$.
Moreover,
\begin{equation}
\label{eq:prime-cubic-resultant}
 B_\ell^3=
 \left|\operatorname{Res}_T\bigl(\Phi_\ell(T),1+T+T^{u+1}\bigr)\right|.
\end{equation}
\end{proposition}

\begin{proof}
The identity $1+u+u^2=0$ in $\mathbb Z/\ell\mathbb Z$ shows that the product
of the three roots is $1$ and that their pairwise products have exponents
$-H$.  This proves the displayed formula for $F_H$.  Proposition~\ref{prop:prime-nonzero}
shows that both internal coefficients are nonzero.  They are algebraic
integers, and $\eta_{-H}$ is a Galois conjugate of $\eta_H$, proving the norm
assertions.

Write $x=\zeta^{u-1}$.  Since $u\ne1\pmod\ell$, $x$ is again primitive, and
\[
 \frac{\eta_H}{\zeta}=1+x+x^{u+1}.
\]
The norm of $\zeta$ is $1$, while $[L:K]=3$.  Taking norms from $L$ to
$\mathbb Q$ and using the resultant formula for the norm of a polynomial in
$\zeta$ gives~\eqref{eq:prime-cubic-resultant}.
\end{proof}

For $\operatorname{ord}_\ell(q)=3$, write $S_4(q,\ell)$ for the assertion that
every monic irreducible factor of $\Phi_\ell$ over $\mathbb F_q$ has Hamming
weight four.

\begin{theorem}
\label{thm:prime-cubic-full-support}
Let $q=p^a$ satisfy $\operatorname{ord}_\ell(q)=3$.  Then
\[
 \boxed{S_4(q,\ell)\quad\Longleftrightarrow\quad p\nmid B_\ell.}
\]
\end{theorem}

\begin{proof}
All irreducible factors have degree three, with root sets the cosets of $H$.
Their characteristic-zero lifts are the Galois conjugates of $F_H$.
If $p\nmid B_\ell$, the two internal coefficients of every such lift are
units at every prime above $p$, so all their reductions have weight four.

Conversely, suppose $p\mid B_\ell$.  Some prime of $K$ above $p$ contains
$\eta_H$.  Extend it to a prime of $L$ and reduce $\zeta$ there.  Since
$p\ne\ell$, its reduction remains a primitive $\ell$th root.  The image of
$H=\langle q\rangle$ is a $q$-Frobenius orbit of length three, so the reduction
of $F_H$ is an irreducible cubic over $\mathbb F_q$.  Its quadratic
coefficient is zero.  Thus $S_4(q,\ell)$ fails.
\end{proof}

\begin{theorem}
\label{thm:prime-cubic-constacyclic-obstruction}
Assume $\operatorname{ord}_\ell(q)=3$ and $p\mid B_\ell$.
For every $\lambda\in\mathbb F_q^\times$, every length-$\ell$
$\lambda$-constacyclic code generated by an irreducible cubic factor of
$X^\ell-\lambda$ has parameters
\[
 \boxed{[\ell,\ell-3,3]_q.}
\]
In particular, all these codes fail to be MDS.
\end{theorem}

\begin{proof}
First take $\lambda=1$.  The preceding theorem supplies an irreducible cubic
factor $g$ with a zero internal coefficient.  The coefficient vector of $g$ is
a nonzero word of its cyclic code and has weight at most three.

Let $\alpha$ be any primitive $\ell$th root in $\mathbb F_{q^3}$.
For the cyclic code with root orbit $\{\alpha,\alpha^q,\alpha^{q^2}\}$, the
evaluation map
\[
 (c_0,\ldots,c_{\ell-1})\longmapsto
 \sum_{i=0}^{\ell-1}c_i\alpha^i
\]
is a parity-check map to the three-dimensional $\mathbb F_q$-space
$\mathbb F_{q^3}$.  Its columns are nonzero.  Two distinct columns cannot be
proportional, because
\[
 \mu_\ell\cap\mathbb F_q^\times=\{1\},
\]
which follows from $\ell\nmid q-1$.  The minimum distance is therefore at
least three.  For the code supplied by $g$, it is exactly three.

The permutations of coordinate positions $i\mapsto bi\pmod\ell$, with
$b\in(\mathbb Z/\ell\mathbb Z)^\times$, act transitively on the root cosets
and yield equivalent cyclic codes.  Thus every irreducible-generator cyclic
code under consideration has minimum distance three.

Finally, $\gcd(\ell,q-1)=1$.  For any $\lambda$ choose
$c\in\mathbb F_q^\times$ with $c^\ell=\lambda$.  The map
\[
 \mathbb F_q[X]/(X^\ell-1)\longrightarrow
 \mathbb F_q[X]/(X^\ell-\lambda),\qquad
 f(X)\longmapsto f(X/c)
\]
is a ring isomorphism and a diagonal equivalence of coefficient spaces.
It carries irreducible cubic generators to their monic scalar multiples
$c^3g(X/c)$ and preserves Hamming weight.  This proves the assertion for
every $\lambda$.  The Singleton bound is $d_H\le4$, so distance three is
not MDS.
\end{proof}

Proposition~\ref{prop:prime-ASB} shows that ASB is four for all the constacyclic constants in this family. Thus the norm obstruction can be strict even when ASB equals Singleton.

\subsubsection{An explicit comparison}
\label{subsec:prime-cubic-examples}

For $\ell=37$ one may take $u=10$.  Exact integer elimination gives
\[
 \operatorname{Res}_T(\Phi_{37}(T),1+T+T^{11})
 =103823=47^3,
 \qquad B_{37}=47.
\]
Moreover, $47\equiv10\pmod{37}$ has order three, and
\[
 g(X)=X^3+5X-1\in\mathbb F_{47}[X]
\]
is an irreducible factor of $\Phi_{37}$.  For example,
$X^{37}\equiv1\pmod{g}$ and $g(1)=5$; every root of $g$ is therefore
primitive of order $37$, whose $47$-Frobenius orbit has length three.
The reciprocal factor is $X^3-5X^2-1$.

There are twelve cubic factors.  Precisely two have weight three and ten
have weight four. Indeed, $47\bmod37\in H$, so $47$ splits completely in $K$; its twelve residue primes correspond to the twelve cubic factors. Moreover,
$v_{47}(B_{37})=1$ means that precisely one residue prime annihilates
$\eta_H$.  Thus precisely one factor has zero quadratic coefficient and
precisely one has zero linear coefficient.  They are distinct, since a
primitive-order-$37$ cubic cannot be $X^3-1$.

For every $q=47^a$ with $3\nmid a$, these twelve factors remain irreducible
and retain the same weight distribution.  For every
$\lambda\in\mathbb F_q^\times$, all the corresponding irreducible-generator
constacyclic codes have parameters $[37,34,3]_q$.  This is an infinite
extension-field family in the fixed characteristic $47$.

Example~\ref{ex:n7} gives the contrasting case $B_7=2$: the norm prime destroys full support but preserves equal weight.

Conversely, $p\nmid B_\ell$ guarantees only that all the irreducible cubic
generators have weight four.  Other codewords can still be relevant to
the minimum distance, so the full-support test alone is not a complete
classification of the MDS property.

\subsection{A quadratic-residue exclusion from a short multiplier}
\label{sec:qr-h5}

This section establishes a characteristic-dependent obstruction for the two
quadratic-residue codes of prime length.  The obstruction is obtained from
two coefficients of an integral quadratic-period polynomial.  Its proof
treats split and inert reduction in the same way and does not require a
quadratic norm to be identified with a finite-field norm.

Let $\ell>7$ be a prime satisfying
\begin{equation}
  \ell\equiv 7 \text{ or }103\pmod {120},
  \qquad r=\frac{\ell-1}{2}.
  \label{eq:qr-h5-length}
\end{equation}
Thus $r$ is odd.  If $\chi$ denotes the quadratic character modulo $\ell$,
then
\begin{equation}
  \chi(-1)=-1,\qquad
  \chi(2)=1,\qquad \chi(3)=\chi(5)=-1.
  \label{eq:qr-h5-characters}
\end{equation}
The oddness of $r$ is essential below; the last three character values alone
do not imply it.

Write $Q$ and $Q'$ for the nonzero quadratic residues and nonresidues in
$\mathbb Z/\ell\mathbb Z$, respectively.  Let $q$ be a prime power of
characteristic $p>5$, and assume
\begin{equation}
  \operatorname{ord}_{\ell}(q)=r.
  \label{eq:qr-h5-order}
\end{equation}
Choose a primitive $\ell$th root $\zeta$ in an algebraic closure of
$\mathbb F_q$, and put
\begin{equation}
  b(X)=\prod_{a\in Q}(1-\zeta^aX)=\sum_{j=0}^{r}b_jX^j,
  \qquad
  \bar b(X)=\prod_{a\in Q'}(1-\zeta^aX)
             =\sum_{j=0}^{r}\bar b_jX^j.
  \label{eq:qr-h5-finite-factors}
\end{equation}
The bar distinguishes the two root sets.  It is not assumed to be an
automorphism of $\mathbb F_q$.
The product indexed by $Q$ has zero set indexed by $-Q=Q'$;
the labels $b$ and $\bar b$ throughout refer to the product indices.

Since the subgroup generated by $q$ modulo $\ell$ is $Q$, both polynomials
belong to $\mathbb F_q[X]$.  Each root set is one Frobenius orbit, so $-b$
and $-\bar b$ are the two monic irreducible factors of degree $r$ of
$\Phi_\ell$.  Moreover,
\begin{equation}
  b_0=\bar b_0=1,\qquad b_r=\bar b_r=-1,
  \qquad b(X)\bar b(X)=\Phi_\ell(X).
  \label{eq:qr-h5-endpoints}
\end{equation}
For example, the leading coefficient follows from
$\sum_{a\in Q}a=0$ in $\mathbb Z/\ell\mathbb Z$ and the oddness of $r$.

\subsubsection{Integral coefficients and the fifth norm difference}
\label{subsec:qr-h5-integral}

Choose a primitive complex $\ell$th root $\xi$ and define
\[
  \mathcal B(X)=\prod_{a\in Q}(1-\xi^aX)
              =\sum_{j=0}^{r}\beta_jX^j.
\]
The coefficients lie in the ring of integers of
$K=\mathbb Q(\tau)$, where
\begin{equation}
  \tau=\sum_{a=1}^{\ell-1}\chi(a)\xi^a,
  \qquad \tau^2=-\ell,
  \qquad \overline\tau=-\tau.
  \label{eq:qr-h5-gauss}
\end{equation}
Indeed, complex conjugation gives $\overline\tau=\chi(-1)\tau=-\tau$,
whereas substitution $a=ub$ yields
\[
  \tau\overline\tau
   =\sum_{u\ne0}\chi(u)\sum_{b\ne0}\xi^{b(u-1)}
   =(\ell-1)-\sum_{u\ne0,1}\chi(u)=\ell.
\]
Here the bar on $K$ is its nontrivial field automorphism.  For
$1\le k\le5$, let $s_k=\sum_{a\in Q}\xi^{ak}$.  The quadratic-period
identity and \eqref{eq:qr-h5-characters} give
\begin{equation}
  s_k=\frac{-1+\chi(k)\tau}{2},\qquad
  s_1=s_2=s_4=\frac{\tau-1}{2},\qquad
  s_3=s_5=\frac{-\tau-1}{2}.
  \label{eq:qr-h5-power-sums}
\end{equation}
Newton's identities, with the signs corresponding to
$\prod(1-\xi^aX)$, are
\begin{equation}
  j\beta_j=-\sum_{k=1}^{j}s_k\beta_{j-k},\qquad \beta_0=1.
  \label{eq:qr-h5-newton}
\end{equation}
These identities are applied over $K$ before reduction.  In the present
calculation their denominators have no prime factors other than $2,3,5$,
so reduction is valid in every characteristic $p>5$.

\begin{lemma}
\label{lem:qr-h5-coefficients}
Under \eqref{eq:qr-h5-length}, the first five integral coefficients satisfy
\begin{align}
  2\beta_1&=1-\tau,\nonumber\\
  8\beta_2&=3-\ell-4\tau,\nonumber\\
  48\beta_3&=15-9\ell+(\ell-7)\tau,\nonumber\\
  384\beta_4&=(\ell-7)(\ell-15+16\tau),\nonumber\\
  3840\beta_5&=-(\ell-7)
           \bigl((\ell-63)\tau-25\ell+135\bigr).
  \label{eq:qr-h5-coefficient-formulas}
\end{align}
\end{lemma}

\begin{proof}
Substitute \eqref{eq:qr-h5-power-sums} successively into
\eqref{eq:qr-h5-newton}.  The first two steps are
\[
  \beta_1=-s_1,\qquad
  2\beta_2=s_1^2-s_2.
\]
Using $\tau^2=-\ell$ gives the first two displayed formulas.  The next
three steps are
\begin{align*}
  3\beta_3&=-s_1\beta_2-s_2\beta_1-s_3,\\
  4\beta_4&=-s_1\beta_3-s_2\beta_2-s_3\beta_1-s_4,\\
  5\beta_5&=-s_1\beta_4-s_2\beta_3-s_3\beta_2
             -s_4\beta_1-s_5.
\end{align*}
Substitution of the preceding rows and another use of $\tau^2=-\ell$
yield the remaining formulas.
\end{proof}

Define the rational integers
\[
  N_j=N_{K/\mathbb Q}(\beta_j)=\beta_j\overline{\beta_j},
\]
and set
\begin{equation}
  F_5(T)=T^3+399T^2-25381T-4275.
  \label{eq:qr-h5-polynomial}
\end{equation}

\begin{proposition}
\label{prop:qr-h5-norm-identity}
The identities
\begin{align}
  N_4&=\frac{(\ell-7)^2(\ell+1)(\ell+225)}{384^2},
  \label{eq:qr-h5-norm-four}\\
  N_5-N_4&=\frac{(\ell-7)^2F_5(\ell)}{3840^2}
  \label{eq:qr-h5-norm-difference}
\end{align}
hold in $\mathbb Q$ and hence as identities between rational expressions
for algebraic-integer norms.
\end{proposition}

\begin{proof}
For rational $A,B$, the norm of $A+B\tau$ is $A^2+\ell B^2$.
Consequently, Lemma~\ref{lem:qr-h5-coefficients} gives
\[
  N_4=\frac{(\ell-7)^2\bigl((\ell-15)^2+256\ell\bigr)}{384^2},
\]
and $(\ell-15)^2+256\ell=(\ell+1)(\ell+225)$.  Similarly,
\[
  N_5=\frac{(\ell-7)^2
       \bigl((25\ell-135)^2+\ell(\ell-63)^2\bigr)}{3840^2}.
\]
The polynomial identity
\[
  (25T-135)^2+T(T-63)^2
      -100\bigl((T-15)^2+256T\bigr)
   =T^3+399T^2-25381T-4275
\]
proves \eqref{eq:qr-h5-norm-difference}.
\end{proof}

Choose a reduction of the integral cyclotomic data at $p$ compatible with
$\xi\mapsto\zeta$.  Then $\beta_j$ and $\overline{\beta_j}$ reduce to
$b_j$ and $\bar b_j$, respectively, and
\begin{equation}
  b_j\bar b_j=N_j\pmod p.
  \label{eq:qr-h5-reduced-norm}
\end{equation}
This product identity is valid whether $p$ splits or remains inert in
$K$.  If its right-hand side is zero, it only asserts that at least one
of the two reduced coefficients is zero.  We shall use precisely this
assertion, without identifying the two reductions.

\subsubsection{The resulting short codeword}
\label{subsec:qr-h5-codeword}

Let
\[
  C_Q=(b),\qquad C_{Q'}=(\bar b)
       \subseteq\mathbb F_q[X]/(X^\ell-1).
\]
Both codes have dimension $\ell-r=r+1$.  Their Singleton distance is
therefore $r+1$.

\begin{theorem}
\label{thm:qr-h5-non-mds}
Assume \eqref{eq:qr-h5-length} and \eqref{eq:qr-h5-order}, with
$p>5$.  If
\begin{equation}
  p\mid F_5(\ell),
  \label{eq:qr-h5-divisibility}
\end{equation}
then $C_Q$ and $C_{Q'}$ are not MDS.  More generally, for every
$\lambda\in\mathbb F_q^\times$, each length-$\ell$ $\lambda$-constacyclic
code whose generator is irreducible of degree $r$ is not MDS.
\end{theorem}

\begin{proof}
Inversion exchanges $Q$ and $Q'$ because $r$ is odd.  Comparing the two
products in \eqref{eq:qr-h5-finite-factors} gives
\begin{equation}
  \bar b(X)=-X^r b(X^{-1}),\qquad
  b_{r-j}=-\bar b_j\quad(0\le j\le r).
  \label{eq:qr-h5-mirror}
\end{equation}
The hypothesis \eqref{eq:qr-h5-divisibility}, together with
\eqref{eq:qr-h5-norm-difference}, implies $N_5=N_4$ modulo $p$.

First suppose $N_4=0$ modulo $p$.  By
\eqref{eq:qr-h5-reduced-norm}, either $b_4=0$ or $\bar b_4=0$.
The corresponding degree-$r$ generator has at most $r$ nonzero
coefficients and is itself a nonzero codeword of weight at most $r$.
That code is not MDS.  To obtain the same conclusion for the other code,
choose any nonsquare $u$ modulo $\ell$.  The map
\[
  c(X)\longmapsto c(X^u)\pmod {X^\ell-1}
\]
is a permutation of the $\ell$ coordinates and exchanges the defining
sets $Q,Q'$.  Thus $C_Q$ and $C_{Q'}$ are monomially equivalent.

Now suppose $N_4\ne0$ modulo $p$.  Then $N_5\ne0$ as well, and all four
coefficients $b_4,\bar b_4,b_5,\bar b_5$ are nonzero.  Consider
\begin{equation}
  c(X)=(b_4-b_5X)b(X)\in C_Q.
  \label{eq:qr-h5-short-word}
\end{equation}
It has degree $r+1<\ell$, is nonzero, and satisfies
\begin{align*}
  [X^5]c&=b_4b_5-b_5b_4=0,\\
  [X^{r-4}]c
    &=b_4b_{r-4}-b_5b_{r-5}
      =-b_4\bar b_4+b_5\bar b_5=0.
\end{align*}
The two positions are distinct.  Indeed,
\eqref{eq:qr-h5-length} with $\ell>7$ implies $\ell\ge103$ and $r\ge51$.
Among the $r+2$ coefficient positions from $0$ to $r+1$, at least two
are zero.  Hence $\operatorname{wt}(c)\le r$, proving that $C_Q$ is not
MDS.  The coordinate equivalence above proves the claim for $C_{Q'}$.

Finally, \eqref{eq:qr-h5-order} implies $\ell\nmid q-1$.  Thus raising
to the $\ell$th power is a bijection of $\mathbb F_q^\times$.  For any
$\lambda\ne0$, choose its unique $\ell$th root
$\delta\in\mathbb F_q^\times$.  Substitution $Y=X/\delta$ induces the
ring isomorphism
\[
  \mathbb F_q[Y]/(Y^\ell-1)
       \longrightarrow\mathbb F_q[X]/(X^\ell-\lambda).
\]
It scales individual coordinates by nonzero field elements and therefore
preserves Hamming weight.  Since
\[
  X^\ell-1=(X-1)(-b(X))(-\bar b(X)),
\]
the two degree-$r$ irreducible generators, after the substitution, give
all degree-$r$ irreducible generators of $X^\ell-\lambda$.  Their codes
are consequently not MDS.
\end{proof}

The argument includes characteristics dividing $N_4$.  In particular,
Theorem~\ref{thm:qr-h5-non-mds} requires neither inert reduction nor
primitivity of $p$ modulo $\ell$.  The condition on $q$ is the one that
makes each quadratic-residue root set a single Frobenius orbit.

\subsubsection{Comparison with three polynomial tests}
\label{subsec:qr-h5-comparison}

For the same character pattern, consider the three polynomials
\begin{align*}
  P_2(T)&=T^3+31T^2-581T-99,\\
  P_4(T)&=T^4+148T^3-7178T^2+22068T-3375,\\
  O_1(T)&=T^2-34T+45.
\end{align*}
Their resultants with \eqref{eq:qr-h5-polynomial}, in the indicated
order, are
\begin{align}
  \operatorname{Res}(F_5,P_2)
    &=2^{27}3^6\cdot113,\nonumber\\
  \operatorname{Res}(F_5,P_4)
    &=-2^{36}3^6 5^2\cdot7\cdot167\cdot1973,\nonumber\\
  \operatorname{Res}(F_5,O_1)
    &=2^{12}3^5\cdot5\cdot2887.
  \label{eq:qr-h5-resultants}
\end{align}
These identities can be checked by the determinant of the Sylvester
matrix.  All four polynomials are monic.  Therefore, if $p>2887$ and
$p\mid F_5(\ell)$, then
\begin{equation}
  P_2(\ell)P_4(\ell)O_1(\ell)\ne0\pmod p.
  \label{eq:qr-h5-nonredundancy}
\end{equation}
Thus the obstruction in Theorem~\ref{thm:qr-h5-non-mds} is disjoint from
the vanishing tests for these three displayed polynomials in that
characteristic range.  This comparison is restricted to this explicit
list of tests.

\subsubsection{A certified parameter}
\label{subsec:qr-h5-certificate}

The hypotheses of Theorem~\ref{thm:qr-h5-non-mds} occur, for example, at
\begin{equation}
  p=81667,\qquad q=p^2,\qquad
  \ell=6669499087=p^2+198.
  \label{eq:qr-h5-example}
\end{equation}
Here $\ell\equiv7\pmod {120}$, and
\begin{equation}
  F_5(198)=18375075=3^2\cdot5^2\cdot81667.
  \label{eq:qr-h5-example-value}
\end{equation}
As $\ell\equiv198\pmod p$, this proves $p\mid F_5(\ell)$.

Appendix~\ref{app:certificate} supplies a complete recursive Lucas certificate for the primality of $p$ and $\ell$, and proves $\ord_\ell(p)=\ell-1$. Therefore $\ord_\ell(q)=r$.

At this parameter, the three comparison polynomials satisfy
\[
  P_2(198)\equiv42543,\qquad
  P_4(198)\equiv60911,\qquad
  O_1(198)\equiv32517\pmod {81667}.
\]
All three residues are nonzero, in agreement with
\eqref{eq:qr-h5-resultants}.  Consequently this certified parameter
exhibits the norm obstruction without invoking any of the three
comparison tests.

Proposition~\ref{prop:prime-ASB} gives ASB equal to $r+1$ for all the
constacyclic constants in Theorem~\ref{thm:qr-h5-non-mds}; hence ASB does not
give this exclusion. The displayed resultant comparison concerns exactly
the three polynomial loci stated above. Other coding-theoretic restrictions
can overlap the broader quadratic-residue family.

For the certified pair $(\ell,p)$, every field $q=p^{2e}$ with
$\gcd(e,r)=1$ has $\ord_\ell(q)=r$, so the theorem gives infinitely many
extension-field parameters. This does not assert infinitely many different
prime lengths.

\section{Further questions}

The finite inverse classification leaves several structural questions. First, one may seek descriptions of the sets $\mathcal A_{n,p}$ in arithmetic terms which avoid the coefficient computations. The symmetries in Section~\ref{sec:symmetry} give sufficient conditions, but need not account for all accidental equalities of weights. Second, bounds for the relevant norm primes can be sharpened using the actual orbit coefficients rather than the universal estimate $2^{\varphi(n)}$. Third, the coefficient obstruction and selected short-multiplier norms may be compared systematically with ASB and with the full maximal-minor obstruction. The quadratic-residue family obtained from a degree-one multiplier illustrates this distinction.

For the explicit quadratic-residue condition, one legitimate characteristic-length pair yields infinitely many extension fields preserving the required Frobenius subgroup. Proving that infinitely many different prime lengths satisfy both the norm divisibility condition and the subgroup-order condition is a separate arithmetic existence problem. No such assertion is needed for the exclusion theorem.

\appendix
\section{A finite certificate for the quadratic-residue family}\label{app:certificate}

The following elementary criterion makes the numerical nonemptiness assertion
independent of a primality-testing black box.

\begin{lemma}\label{lem:Lucas}
Let $M>2$ and suppose the complete prime factorization of $M-1$ is known.
If an integer $a$ satisfies
\[
 a^{M-1}\equiv1\pmod M,\qquad
 \gcd\bigl(a^{(M-1)/s}-1,M\bigr)=1
 \quad\text{for every prime }s\mid M-1,
\]
then $M$ is prime and $a$ has order $M-1$ modulo $M$.
\end{lemma}
\begin{proof}
For any prime $v\mid M$, the order of $a$ modulo $v$ divides $M-1$.
The gcd conditions force that order to contain the full power of each prime
dividing $M-1$, so it equals $M-1$. Hence $M-1\mid v-1$, which forces
$v\ge M$. Thus $M=v$ is prime, and the same order conclusion holds modulo $M$.
\end{proof}

For each row of Table~\ref{tab:Lucas}, direct modular exponentiation gives
$a^{M-1}\equiv1\pmod M$. The last column lists
$a^{(M-1)/s}\bmod M$ in increasing order of the distinct primes $s$
in the second column. Each listed value $v$ satisfies $\gcd(v-1,M)=1$.
Beginning with the prime $2$, every prime divisor required by a later row
has therefore already been certified by an earlier row. The last row also
certifies the required primitive-root condition.

\begin{center}
\small
\setlength{\tabcolsep}{4pt}
\begin{longtable}{@{}rllp{76mm}@{}}
\caption{Recursive Lucas certificate.}\label{tab:Lucas}\\
\toprule
$M$ & $M-1$ & $a$ & Residues for the distinct prime divisors\\
\midrule
\endfirsthead
\toprule
$M$ & $M-1$ & $a$ & Residues for the distinct prime divisors\\
\midrule
\endhead
$3$ & $2$ & $2$ & $2$ \\
$5$ & $2^{2}$ & $2$ & $4$ \\
$7$ & $2\cdot3$ & $3$ & $6, 2$ \\
$11$ & $2\cdot5$ & $2$ & $10, 4$ \\
$13$ & $2^{2}\cdot3$ & $2$ & $12, 3$ \\
$17$ & $2^{4}$ & $3$ & $16$ \\
$23$ & $2\cdot11$ & $5$ & $22, 2$ \\
$29$ & $2^{2}\cdot7$ & $2$ & $28, 16$ \\
$47$ & $2\cdot23$ & $5$ & $46, 25$ \\
$349$ & $2^{2}\cdot3\cdot29$ & $2$ & $348, 226, 257$ \\
$9871$ & $2\cdot3\cdot5\cdot7\cdot47$ & $3$ & $9870, 9219, 4996, 5180, 2759$ \\
$37537$ & $2^{5}\cdot3\cdot17\cdot23$ & $13$ & $37536, 25417, 26651, 11201$ \\
$81667$ & $2\cdot3^{2}\cdot13\cdot349$ & $3$ & $81666, 38354, 29309, 78230$ \\
$6669499087$ & $2\cdot3^{2}\cdot9871\cdot37537$ & $81667$ & $6669499086, 5846055168, 2150963284, 6574342481$ \\
\bottomrule
\end{longtable}
\end{center}

For $p=81667$ and $\ell=6669499087$, the resulting order data are
\[
 \ord_\ell(p)=\ell-1,\qquad
 \ord_\ell(p^2)=\frac{\ell-1}{2}=3334749543.
\]
The equality $\ell=p^2+198$ and the integer factorization
$F_5(198)=3^2\cdot5^2\cdot81667$ prove the norm divisibility condition.
The example is used only to certify nonemptiness of the general theorem.

\end{document}